\documentclass[a4paper,10pt,twocolumn]{revtex4}
\usepackage[utf8]{inputenc}
\usepackage{amsmath}
\usepackage{amssymb}
\usepackage[amssymb, thinqspace]{SIunits}
\usepackage[english]{babel}
\usepackage{graphicx}
\usepackage{natbib}
\usepackage{bm}
\usepackage{xcolor}
\newcommand{\diff}[2][]{\frac{\mathrm{d} #1}{\mathrm{d} #2}}

\renewcommand{\vec}[1]{\boldsymbol{#1}}
\begin{document}

\author{C.~C.~Maa\ss}
\author{N.~Isert}
\author{G.~Maret}
\author{C.~M.~Aegerter}
\affiliation{Physics Department, University of Konstanz, Box M621,
78457 Konstanz, Germany}
\title{Experimental investigation of the freely cooling granular gas}

\begin{abstract}
Using diamagnetically levitated particles we investigate the dynamics of the freely cooling granular gas. At early times we find good agreement with Haff's law, where the time scale for particle collisions can be determined from independent measurements. At late times, clustering of particles occurs. This can be included in a Haff-like description taking into account the decreasing number of free particles. With this a good description of the data is possible over the whole time range.
\end{abstract}
\maketitle

In the study of granular media it has proven useful to define states on the basis of the solid, liquid and gaseous states of a molecular substance~\cite{Jaeger1996}. In this vein of definition, a granular gas is a dilute system of macroscopic particles in random, quasi-Brownian motion. In contrast to a molecular gas, collisions between granular particles are only partially elastic. Hence, energy has to be constantly injected into the system and one expects the manner of excitation to have a non-negligible effect on the state of the system \cite{vanZon2004c}. However, the independent characteristics of the granular gas should be governed by the statistical properties of collisions and by energy losses due to inelasticity.
As the behaviour of an excited granular gas can be divided into short-time incipient cooling states between single excitation events, an investigation of the cooling process allows us to study fundamental characteristics of the granular gas independent from the specific manner of excitation \cite{Santos1989,Garzo1999,Montanero2000}.

Haff\cite{Haff1983} derived a hydrodynamic theory of granular
motion. The resulting cooling law states that the
kinetic energy of a spatially isotropic granular gas without
external driving should decrease like $1/t$. This behaviour emerges
after a characteristic time determined by the density, initial speed
and particle properties.

An experimental study of the cooling of a 2-d granular gas on a surface together with a comparative MD simulation has been reported in \cite{Painter2003}, with a special emphasis on clustering behaviour. Haff's law was not observed, which might have been due to additional energy loss by surface friction.

To extend Haff's picture, the freely cooling granular gas has been investigated in several analytical and simulation studies. Here the main issue is to describe the process in the presence of inhomogeneities, presumably introduced by inelastic collapse. The authors incorporated  clustering and inelastic collapse \cite{Ben-Naim1999,Nie2002} as well as a velocity-dependent restitution coefficient \cite{Poeschel2003,Ramirez1999}. Although all studies agree in a decrease of cooling with time, the resulting cooling exponents are still a matter of debate. In all studies the quantity studied is the granular temperature $\langle v^2\rangle$, which is, however, increasingly ill-defined for clustering states. Furthermore, as we will see below in a simple derivation of Haff's law, the speed of the particles is a more fundamental quantity. Also, the cooling process mainly involves the non-clustered particles, and in a video microscopy setup with a limited spatial resolution it is practically impossible to determine any very small particle speeds in the clustered parts of the sample.
Therefore, the observed quantity in our study is the mean particle speed outside of a possible cluster.

\setlength{\unitlength}{1cm}
\begin{figure}[htp]
\includegraphics[width=.42\textwidth]{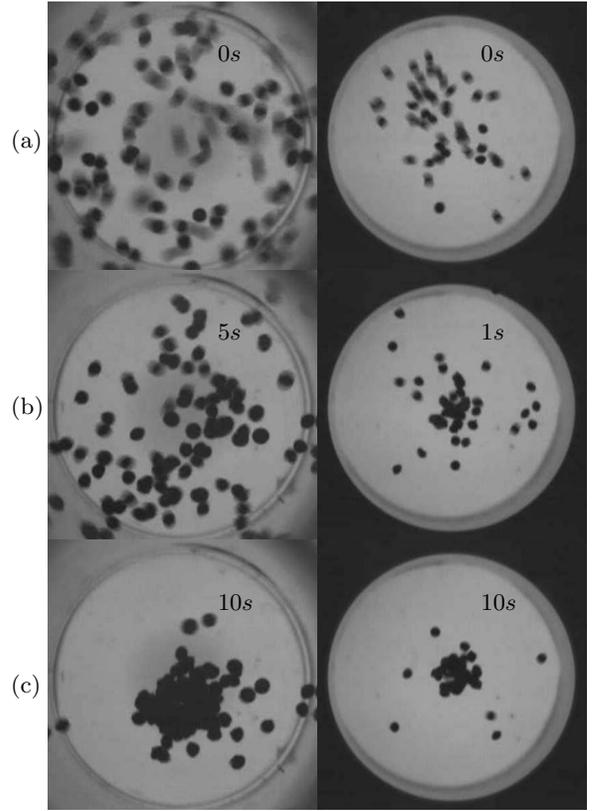}
\begin{picture}(10,0)
\put(0.25,9.35){(a)}
\put(0.25,5.8){(b)}
\put(0.25,2.1){(c)}
 \put(3,10.5){\textcolor{black}{$0s$}}
\put(3,6.8){\textcolor{black}{$5s$}}
\put(3,3.2){\textcolor{black}{$10s$}}
 \put(6.5,10.5){\textcolor{black}{$0s$}}
\put(6.5,6.8){\textcolor{black}{$1s$}}
\put(6.5,3.2){\textcolor{black}{$10s$}}

\end{picture}

 \caption{Cooling process for mechanically (left) and magnetically (right) excited system. Snapshots taken at the driving switch-off (a), the Haff time (b) and after cooling has completed (c).}\label{fig:cooling}
\end{figure}

We will briefly discuss the physics behind Haff's law.  Assume a granular gas in motion, but with no external forcing.
If the density and momentum are spatially isotropic, the time evolution of the energy $T$, which is purely kinetic, can be described by the loss due to inelastic collisions taking place at the rate $\tau^{-1}$:
\begin{equation}\label{eq:energy}\diff{t}{T}=-\frac{(1-\varrho)T}{\tau} \end{equation}
Here we define the restitution coefficient $\rho$ as the amount of kinetic energy retained after a collision. The collision time $\tau$ is given by $s/v$, where $v$ is the speed
and $s$ the  mean free path. In the dilute limit, $s$ is given by
the density $n_0$ and the cross section of the particles $\sigma$,
$s=1/(n_0\sigma)$. As $T \propto v^2$, equation (\ref{eq:energy}) can
be transformed to one describing $v$, the integration of which
yields
\begin{align}\label{eq:haff} \langle v(t) \rangle &\propto \frac{v_0}{1+t/\tau_H} & \tau_H &=\frac{2}{v_0(1-\varrho)n_0\sigma},\end{align}
which is Haff's law in its original form, where the Haff time $\tau_H$ specifies  the system relaxation time scale and depends only on the restitution coefficient $\varrho$ and the particle size.

\begin{figure}[t!p]
 \includegraphics[width=.48\textwidth]{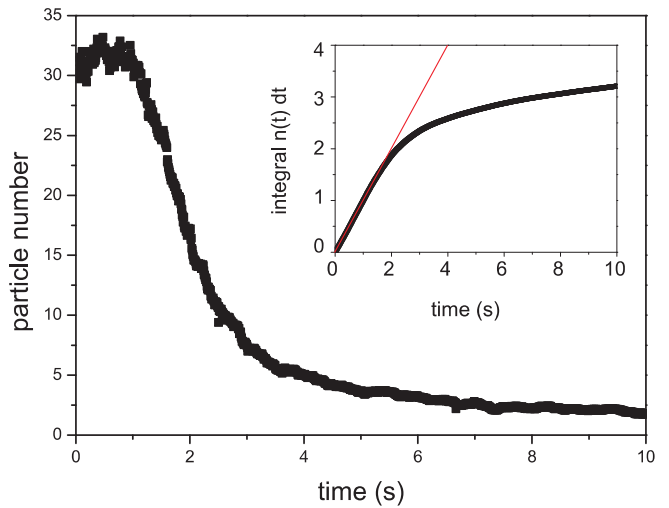}
\caption{Number of particles outside the cluster vs. time}\label{fig:pnum}
\end{figure}

As will be discussed below, the assumption of uniformity does not hold for a system cooled down to a clustering state, which is of course spatially correlated. Thus the mean free path of particles outside the cluster increases and we expect the cooling exponent to deviate from $-1$ for large times and highly clustered states \cite{Goldhirsch1993,McNamara1994,Poeschel2002}. For this, we assume a collision time $\tau=1/(n(t)\sigma v) $ depending on the time dependent number density of particles outside the cluster, $n(t)=n_0f(t)$. Inserting this into (\ref{eq:energy}) and integrating leads to:
\begin{equation}\label{eq:newhaff} v(t)=\frac{v_0}{1+\frac{1}{\tau_H}\int_0^t\;f(t)\mathrm{d}t} \end{equation}
 Note that for constant particle numbers this is equivalent to Haff's original law.
$f(t)$ can be determined experimentally (see Fig. \ref{fig:pnum}) and  integrated numerically from the measured data, such that a parameterless fit to the cooling data is possible.

The above description will hold until the mean free path has grown to the size of the container. At this point, which roughly corresponds to the state where only few free particles are left, we will return to a Haff-like behaviour, where the container size has to be used as the mean free path and the restitution coefficient of collisions with the walls applies.

In order to observe the cooling process experimentally in three dimensions, we have to compensate gravity, otherwise the collision statistics would be dominated by collisions with the bottom of the sample cell after an excitation switch-off~\cite{Ojha2004}. While this compensation is possible in principle with parabola flights or satellites \cite{Falcon1999}, it can be accessed more easily by the use of diamagnetic levitation, which is highly controllable and reproducible. Diamagnetic levitation has previously been applied successfully in studies of granular demixing \cite{catherall2005}, but we are not aware of applications in the field of fundamental properties like velocity distributions and cooling behaviour.

A diamagnetic particle in an external magnetic field $\vec{B}$ acquires an induced magnetic moment antiparallel and  proportional to $\vec{B}$ and thereby a potential energy $U\propto B^2$. Accordingly, in an inhomogeneous field, the force acting on the material is determined by the product of field and field gradient, $\vec{F}\propto\vec{B}\cdot\nabla\vec{B}$, as can be easily verified by taking the spatial derivative of the magnetic energy.
If the resulting $z$-acceleration compensates gravity,
\[ a_z = \frac{\chi}{\mu_0\rho}B\cdot\partial_zB\stackrel{!}{=}g,\]
the particle can be held in a stable state of suspension comparable to actual weightlessness \cite{Berry1997, Braunbek1939}.

\begin{figure}[th]
\includegraphics[width=.5\textwidth]{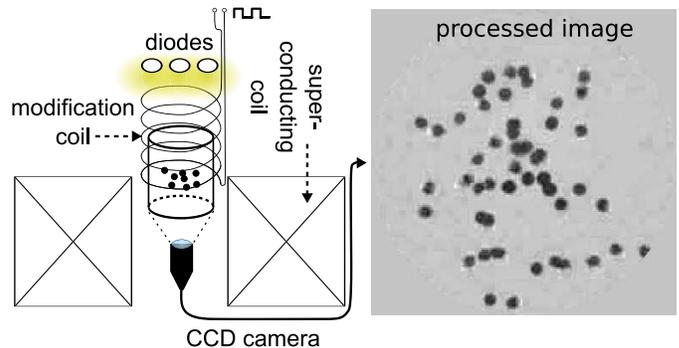}
\caption{Setup for magnetic shaking, including sample cell, field modification coil, diffuse illumination and CCD video observation.  }\label{fig:setup}
\end{figure}

\begin{figure}[th]
 \includegraphics[width=0.5\textwidth, viewport= 0 0 310 210]{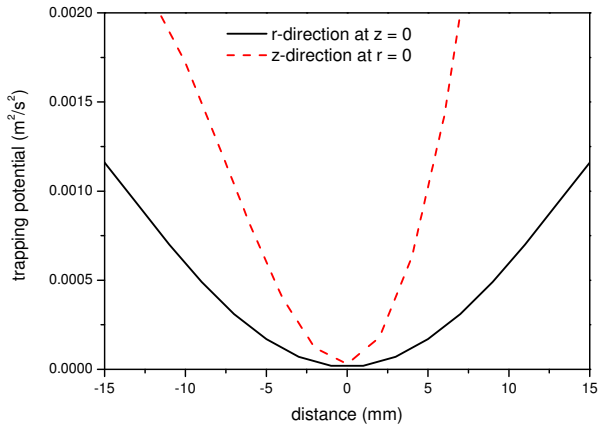}
\caption{Potential maps (gravitational potential subtracted) in $z$ and $r$ for the magnet used in our experiment.  }\label{fig:potential}
\end{figure}

Our setup consists of a helium cooled superconducting magnet coil (Nb$_3$Sn, Oxford) with a cylindrical room temperature bore of \unit{4}{\centi\metre} diameter which is accessible for experiments. The coil can generate field strengths up to \unit{20}{\tesla}, which corresponds to a maximum $B_z\cdot\partial_z B_z$ of approx. \unit{1800}{\square\tesla/\metre}.
Fig.~\ref{fig:potential} shows the potential shapes calculated from the magnetic field chart supplied by the coil manufacturers.  As can be seen, the residual acceleration in a volume around \unit{1}{\centi\metre} at the levitation point ($r,z=0$) are of the order of a few $10^{-3}g$, so that we can effectively speak of milligravity. The  radial potential is bowl-shaped, the height can be tuned by changing the maximum coil field $B_0$ for a given type of diamagnetic material. It is not possible to flatten the radial potential entirely.

The levitation point is determined by field strength $B_0$ and geometry. Hence it is possible to shift this point periodically by modulating the field by an amount $\Delta B$ using an additional coil, effectively shaking the particles magnetically. The resulting acceleration is proportional to $B_0\Delta B$ and acts on all particles in the sample in a similar manner, with variations due to particle height and field inhomogeneities. This spatially almost uniform excitation provides a certain amount of chaos and eventually leads to a decrease in collective sample movement in favour of random single particle motion. Any remaining collective motion can be compensated by choosing a centre--of--gravity reference frame.

We also use the common method of exciting a granular gas, which is to put the sample on top of a mechanical shaker, e.~g. a loudspeaker. In the case of a levitated sample, the moving piston barely touches the lower end of the sample. Particles moving downwards to the topmost piston position are re-injected upwards with a comparatively high energy, which is dissipated by chaotic collisions with the bulk particles.
It is useful to have access to two fundamentally different states of excitation: we are able to examine how far the excitation method characterises the continuously driven gas and to what extent different initial states influence the cooling process.

To get rid of air friction and hydrodynamic interactions as well as to avoid water condensation on the particle surfaces, the sample cells are built to hold a weak vacuum for several hours after depressurisation.

A schematic of the setup for the magnetically shaken case is depicted in Fig.~\ref{fig:setup}. The sample cell is positioned at the point of maximum field gradient located near the top end of the superconducting coil. The small copper modification coil is driven with a square wave current from a computer controlled AC power supply. The setup is illuminated from above by a diode array and a Teflon diffuse layer and observed with a CCD camera from below. In the case of the mechanical excitation, a plastic rod connects the sample cell downwards to a commercial speaker membrane driven by the power supply mentioned above. In this case the setup is illuminated from below and observed from above.

The system's $xy$-projection is observed with a lipstick camera and further processed frame by frame by a blob recognition software (RSI IDL). Thus, we record velocity snapshots by measuring the distance between next neighbours on consecutive frames at a constant frame rate of 120 fps. Although there is a risk of misidentification and thus a systematic underestimation of the velocities of fast particles, this method is reasonably correct for the relatively small speeds relevant in cooling processes.

In the magnetically shaken case, our sample consists of approx.~50 round Bismuth shots  weighing \unit{2\pm 0.1}{\milli\gram} each; in the speaker setup the particle number is approx.~90. Bismuth, which is highly diamagnetic, requires a field of \unit{13.7}{\tesla} (with $B_z\cdot\partial_z B \approx\unit{1000}{\square\tesla/\metre}$) to levitate. The metallic nature of the sample prevents electrostatic charges, so there is no additional Coulomb interaction. However, Bismuth has an extremely low restitution coefficient of approx. 0.1-0.2, so we are operating in the limit of a highly dissipative granular gas.

The energy input from the copper coil is very small, so that the magnetically excited system will only produce reasonable particle speeds if the shaking is done at a resonance frequency determined by the system geometry and the potential shape. Thus, both systems are excited at a non-varying frequency of \unit{1.57}{\hertz}. In a measurement series, the system reaches an excited steady state over a heating period of \unit{10}{\second} in the mechanically and \unit{20}{\second} in the magnetically shaken case, after which the speaker or coil is switched off. Video recording starts precisely at the onset of shaking and is terminated manually when the system has relaxed into a clustered state.
Each experimental series consists of 50 such movies, the well-defined starting time of the movie allows for an easy determination of the onset of cooling $t_0$ in each case, so we do not expect significant systematic errors in averaging over the whole series.

\begin{figure*}
\includegraphics[width=.48\textwidth]{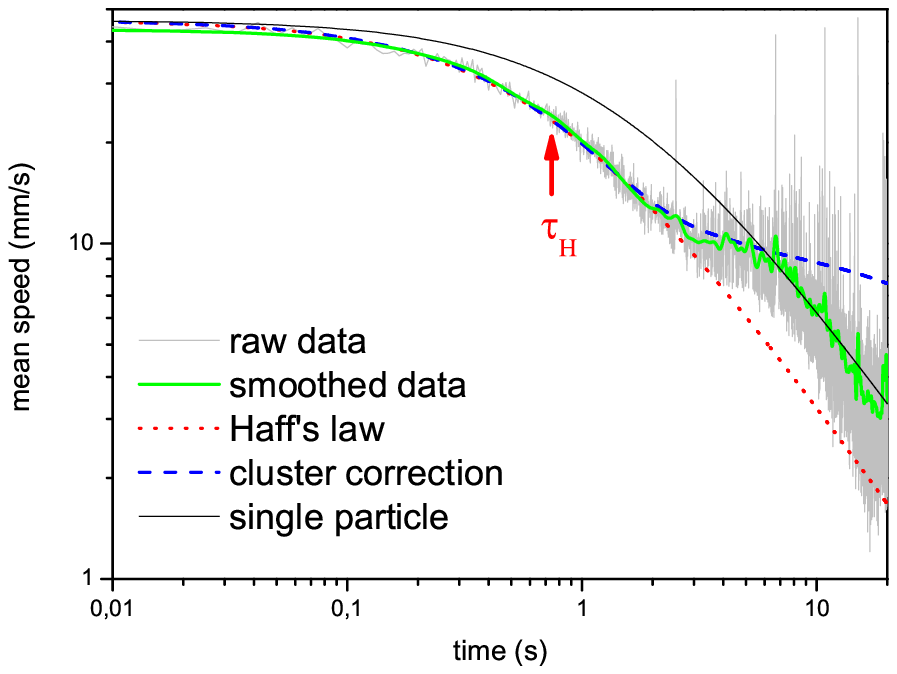}\includegraphics[width=.48\textwidth]{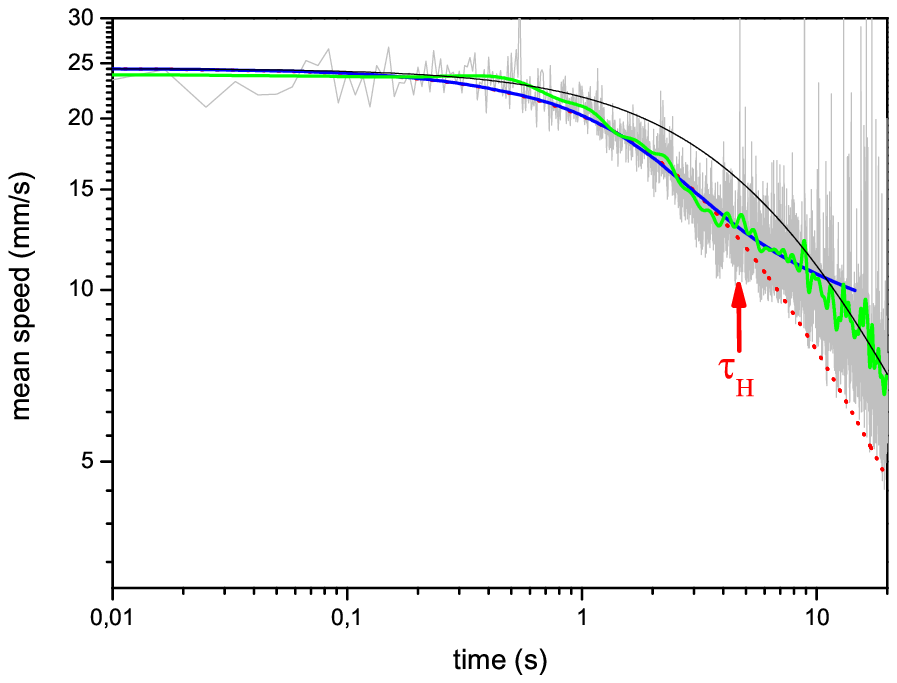}
\caption{Mean particle velocity vs. time for the mechanically (left) and magnetically (right) excited system.  Haff's law without (solid, see Eq.~\ref{eq:haff}) and with (dashed, see Eq.~\ref{eq:newhaff}) cluster correction. Time starts from end of shaking, arrows mark the \mbox{Haff time $\tau_H$}}\label{fig:haff}
\end{figure*}

In Fig.~\ref{fig:haff} we have plotted the mean speed for the remaining free particles outside the cluster. The experimental data averaged over 50 experiments are shown in light grey, the darker line includes an additional smoothing in time over 50 frames each.  In all plots, dotted lines delineate Haff's law (Eq.~\ref{eq:haff}) and dashed lines the corrections for time-dependent particle numbers (Eq.~\ref{eq:newhaff}). All parameters are known from independent experiments: the excited state yields $v_0$, while the container size, particle number $n_0$ and size $\sigma$ are fixed. Thus there are no adjustable parameters in the curves. The cooling behaviour starts to deviate from Haff's law as soon as the cluster begins to form. This can happen at later times than  the typical Haff time, as can be seen in the mechanically shaken sample. In the magnetically shaken sample, this roughly coincides with the Haff time, which is marked by an arrow in Fig.~\ref{fig:haff}. See also the snapshots in Fig.~\ref{fig:cooling}. The deviations are in very good agreement with our calculations derived from the extracted particle numbers. Note also that we have no free fitting parameters left, as speed and mean free path can be extracted from the video data and the restitution coefficient can be established independently.

Especially in the mechanically excited system, we can observe that at a time of about 10 seconds after switch-off, when all but a few particles have merged with the cluster (Fig.~\ref{fig:pnum}), the system seems to revert to a Haff-like behaviour.

We see that there are no qualitative differences in the plots for different excitation methods. This implies that both methods produce a similar state of initial excitation. However, the experiments differ in sample density and particle speeds, which allows us to probe the parameter space  and test the scaling of Haff's law with respect to these parameters.

In conclusion, we have demonstrated a novel method to study granular cooling directly, validating Haff's law. The data can be described by the decreasing density of the free particles outside the cluster. This gives a description of the cooling ground state of a granular gas even in the presence of clustering. Theoretically, the challenge remains to calculate the temporal dependence of the density of free particles from clustering properties in order to have a full description of the system.

We gratefully acknowledge funding by the Deutsche
Forschungsgemeinschaft, the IRTG 667 and the Landesstiftung
Baden-W\"urttemberg.

\end{document}